\documentclass{article}
\usepackage{spconf}
\usepackage{amsmath,graphicx}
\usepackage[hidelinks]{hyperref}  
\usepackage{standalone}
\usepackage{booktabs}
\usepackage{multirow}
\usepackage{makecell}
\usepackage{subcaption}
\usepackage{adjustbox}
\usepackage{hyperref}
\usepackage{enumitem}

\title{Frame-Stacked Local Transformers for Efficient\\Multi-Codebook Speech Generation}
\name{
\begin{tabular}{c}
Roy Fejgin\thanks{Correspondence: rfejgin@nvidia.com, pneekhara@nvidia.com}, Paarth Neekhara, Xuesong Yang, Edresson Casanova, Ryan Langman, \\ Jaehyeon Kim, Subhankar Ghosh, Shehzeen Hussain, Jason Li
\end{tabular}
}

\address{NVIDIA}
\begin{document}
%
\maketitle
\begin{abstract}
Speech generation models based on large language models (LLMs) typically operate on discrete acoustic codes, which differ fundamentally from text tokens due to their multi-codebook structure. At each timestep, models must predict $N$ codebook entries jointly, introducing dependencies that challenge simple parallel prediction approaches. Parallel prediction assumes independence among codebooks, yielding efficient decoding but often at the cost of reduced fidelity. To address this, hierarchical strategies employ a local transformer (LT) to refine predictions and capture intra-timestep dependencies. In this work, we systematically investigate two LT architectures: an autoregressive transformer that generates codebooks sequentially, and a MaskGIT-based transformer that performs iterative masked prediction. Both designs further enable frame stacking, where the primary transformer predicts multiple frames jointly, and the LT decodes their codebooks, offering improvements in speed without compromising perceptual quality. Through extensive analysis, we characterize the tradeoffs between parallel and iterative sampling strategies across different throughput and quality regimes. Finally, we propose practical guidelines for selecting decoding strategies based on deployment priorities such as computational efficiency and synthesis fidelity\footnote{Audio demo page: {\scriptsize \url{https://frame-stacking-lt.github.io}}}
.
\end{abstract}
\section{Introduction}
\label{sec:intro}

Recent advances in large language models (LLMs) for speech generation, such as text-to-speech (TTS) systems, have enabled highly natural speech synthesis by predicting sequences of discrete acoustic codes~\cite{chen2024vall,neekhara2024improving,hussain2025koel,yanguniaudio,wang2024speechx}. Unlike text generation, which produces a one-dimensional sequence of tokens, acoustic representations are typically structured as a matrix of shape $[T, N]$, where $T$ denotes the number of timesteps and $N$ the number of codebooks. Therefore, each timestep requires the prediction of multiple codebook entries. This structural difference introduces challenges unique to speech generation: while text tokens are conditionally dependent only along the temporal dimension, acoustic tokens also exhibit inter-codebook dependencies within each timestep.   

A common baseline for this problem is \emph{parallel prediction}, in which all $N$ codebooks of a timestep are predicted simultaneously under an independence assumption~\cite{neekhara2024improving,hussain2025koel}. While computationally efficient, this approach neglects the rich dependencies among the codebooks, often leading to a degradation in synthesis quality. To mitigate this limitation, alternative decoding strategies introduce \emph{hierarchical modeling} through an auxiliary local transformer (LT), tasked with capturing intra-timestep dependencies~\cite{yanguniaudio,defossez2024moshi}. In such approaches, a primary transformer predicts coarse acoustic embeddings across timesteps, and a secondary LT refines the prediction by modeling the structured codebooks corresponding to each frame. 
This hierarchical transformer setup with an LT has also been successful in diffusion-based autoregressive models that predict continuous codec representations instead of discrete codes~\cite{peng2025vibevoice}.
Note that unlike some alternate approaches with multiple transformers~\cite{chen2024vall,wang2024speechx}, the LT operates within a single frame of acoustic codes instead of multiple frames. 
Alternatively, the delay pattern~\cite{musicgen} partly addresses the issues with parallel prediction. However, it does not capture the complete dependencies amongst the tokens~\cite{yanguniaudio}.

This work examines two principal design choices for the local transformer: (i) an \emph{autoregressive (AR) LT}, which generates codebooks sequentially by conditioning each prediction on previously generated codebooks, and (ii) a \emph{MaskGIT-inspired LT}, which employs iterative masked prediction to jointly model codebooks. Both methods allow for the integration of \emph{frame stacking}, where the primary transformer predicts multiple consecutive frames, and the LT resolves the corresponding codebooks. This hierarchical strategy offers the potential to substantially increase generation speed while maintaining perceptual quality, striking a balance between throughput and fidelity.  
In this study, we systematically analyze the tradeoffs of parallel versus iterative sampling strategies for acoustic code prediction in speech generation LLMs.
\textbf{Our contributions are as follows:}

\begin{itemize}\setlength\itemsep{0em}
    \item We study two iterative multi-codebook prediction heads---autoregressive and MaskGIT---for LLM-based speech synthesis models. We demonstrate that both of these techniques outperform a parallel prediction head in terms of audio quality and target speaker similarity for zero-shot TTS models.
    \item When combined with frame stacking, we demonstrate that a model equipped with an AR LT can achieve 2.1x faster throughput while improving Fréchet Distances and preserving intelligibility and speaker similarity and neural MOS compared to a parallel prediction baseline without frame stacking. This frame-stacking approach allows us to significantly improve throughput of the TTS model without having to retrain a lower frame rate speech codec.
    \item We distill practical guidelines for selecting among these approaches, offering insights into how design choices can be aligned with deployment requirements such as real-time synthesis or offline batch generation.
\end{itemize}

\section{Methodology}
\label{sec:methodology}

\subsection{Background: Multi-codebook Neural Audio Codecs}
Neural audio codecs represent audio as a sequence of discrete tokens arranged in a two-dimensional structure of size $(T, N)$, where $T$ is the number of temporal frames and $N$ is the number of codebooks per frame. Each audio frame is thus described by a set of $N$ codebook entries.  
A key challenge arises because these codebooks are not statistically independent. For Residual Vector Quantization (RVQ)~\cite{rvq} this is true by construction, since higher codebooks are constructed from the residual of lower ones. But even for Finite Scalar Quantization (FSQ)~\cite{fsq} there is nothing in the training objective directly enforcing independence between the codebooks. 
Conventional parallel prediction strategies assume independence among the $N$ codebooks, enabling efficient decoding but often producing artifacts due to mismatches across codebooks~\cite{neekhara2024improving,hussain2025koel}. This motivates the use of iterative prediction strategies, where codebooks within a frame are decoded in a dependent manner, better capturing the intra-frame structure.

\subsection{Base Model: Koel-TTS}
\begin{figure}[t] 
  \centering
  \includegraphics[width=0.8\columnwidth]{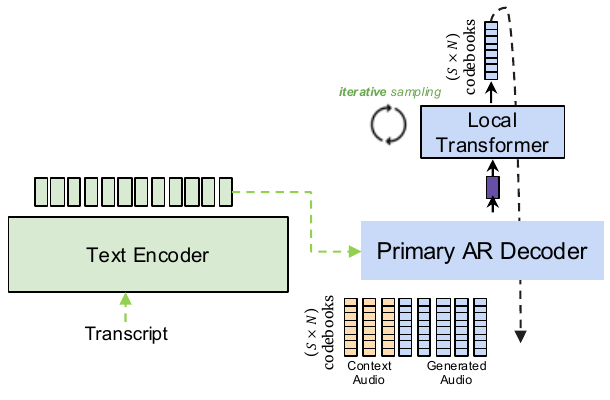}
  \caption{Model Architecture}
  \label{fig:lt_arch}
  \vspace{-5mm}
\end{figure}

Our study builds on Koel-TTS, an encoder-decoder TTS system~\cite{hussain2025koel}. The encoder processes linguistic inputs, while the decoder autoregressively predicts discrete acoustic tokens (Figure~\ref{fig:lt_arch}). For audio tokenization we use a 21.5 frames per second, 8-codebook NanoCodec~\cite{casanova2025nanocodec}. Koel-TTS predicts all $N$ codebooks at each timestep in parallel. While efficient, this parallel decoding inherits the limitations discussed above, that is, the intra-frame dependencies are ignored. 
Consequently, the generated speech suffers from reduced fidelity, which we empirically demonstrate in our experiments (Section~\ref{sec:experiments}).  
To address this, we augment Koel-TTS with an LT that explicitly models intra-frame dependencies, enabling iterative prediction of the codebooks.

\subsection{Local Transformer for Iterative Prediction}
The LT operates at the frame level and refines the predictions of the primary decoder. Given the hidden state output by the primary decoder at a given frame, the LT predicts codebooks iteratively, either autoregressively or through MaskGIT.

\vspace{-3mm}
\subsubsection{Autoregressive Local Transformer}
In the autoregressive variant, the LT generates the $N$ codebook entries sequentially within each frame. Conditioned on previously generated codebooks and the hidden state from the primary decoder, the LT predicts the next codebook until all $N$ entries are decoded. This formulation assumes causal dependencies from codebooks $1$ to $N$ which is true by construction for RVQs. However, even for FSQs, we find that this approach yields better results than parallel prediction, albeit at the cost of increased latency proportional to $N$.

\vspace{-3mm}
\subsubsection{MaskGIT Local Transformer}
In the MaskGIT~\cite{chang2022maskgit} variant, the LT starts from a fully masked sequence and progressively unmasks it through iterative prediction. The LT is configured with non-causal self-attention. The input sequence is initialized with the hidden state followed by $N$ mask embeddings. After each forward pass, a subset of predictions is retained and unmasked, and this partially unmasked sequence is fed into the next iteration. After 
$P$ iterations, all $N$ codebooks are decoded. Importantly, $P$ can be smaller than $N$, allowing multiple tokens to be predicted in parallel, thereby speeding up inference and allowing a flexible tradeoff between speed and quality. In addition to its speed benefits, this  scheme also enables \emph{bidirectional} dependency modeling between codebooks in the frame, without constraining decoding to a particular codebook order (unlike the AR LT). The main drawback is that dependencies between codebooks that get unmasked in the same iteration are not captured, i.e. the modeling of the joint probability distribution is \emph{approximate}, as discussed in~\cite{besnier2025halton}.
\subsection{Frame Stacking}
The hierarchical setup of the primary decoder and LT gives us an opportunity to offload work to the LT and let the primary decoder operate at a lower frame rate, further improving efficiency. We call this \emph{frame stacking}. Concretely, the primary decoder is trained to predict $S$ frames, i.e. $S\times N$ codebooks, in one step. We pass the hidden state from the primary decoder into the LT. The LT then predicts all $S\times N$ codebooks. We study both variants of the LT combined with frame-stacking. We use separate embedding tables for codebooks at different frame indices within the frame stack to disambiguate between them; these embeddings are shared between the primary decoder and the LT. At the input to the primary decoder they are averaged across both the frame stack and codebooks. The frame-stacking setup is faster than the baseline for two reasons. First, the LT is much \emph{smaller} than the primary decoder. Second, \emph{it operates on a short sequence} of up to ~$N+1$ elements, compared to the primary decoder which must self-attend to the entire generation history and also cross-attend to the text encoder. These speed benefits increase with the stacking factor $S$, as explored in section \ref{sec:experiments}.

\vspace{-4mm}
\section{Experiments}
\label{sec:experiments}

\begin{figure*}[t]
  \centering
  \begin{subfigure}{0.24\textwidth}
    \includegraphics[width=\linewidth]{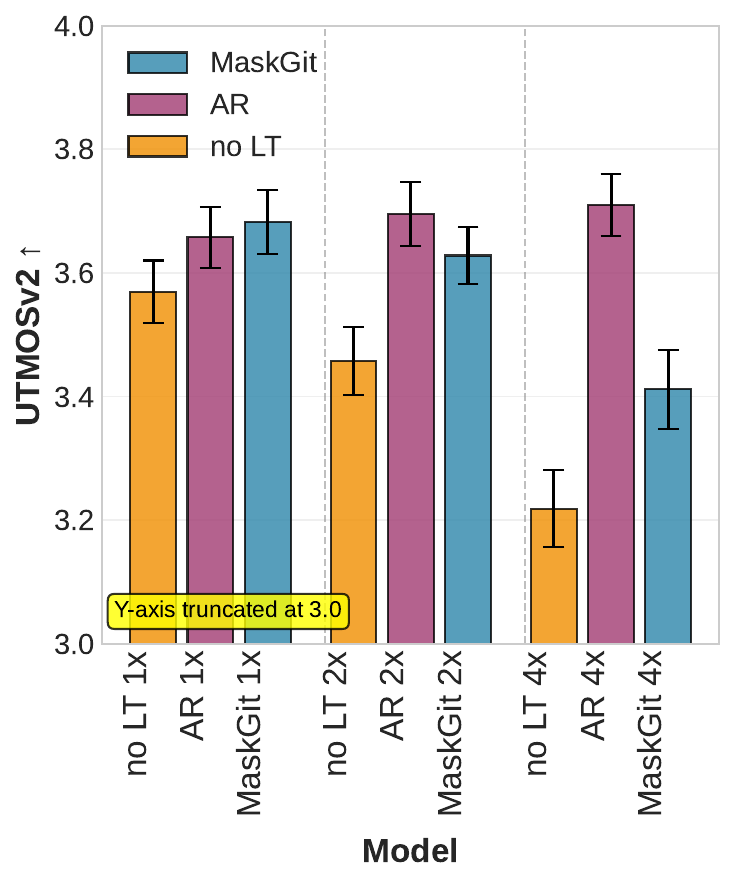}
    \caption{UTMOSv2 (unseen speakers)}
    \label{fig:1a}
  \end{subfigure}
  \hfill
  \begin{subfigure}{0.24\textwidth}
    \includegraphics[width=\linewidth]{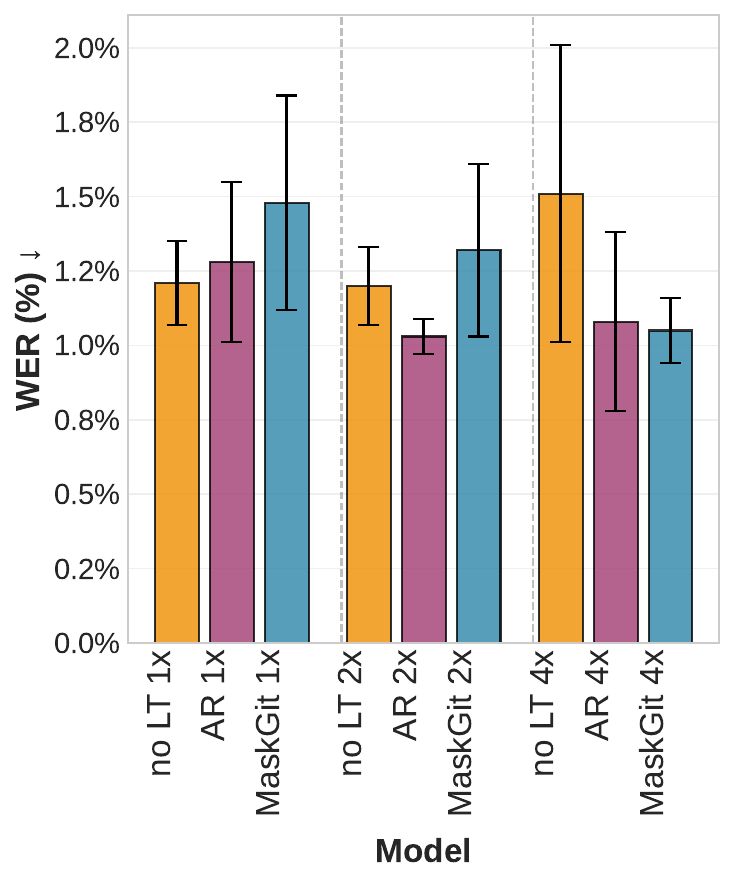}
    \caption{WER (unseen speakers)}
    \label{fig:1b}
  \end{subfigure}
  \hfill
  \begin{subfigure}{0.24\textwidth}
    \includegraphics[width=\linewidth]{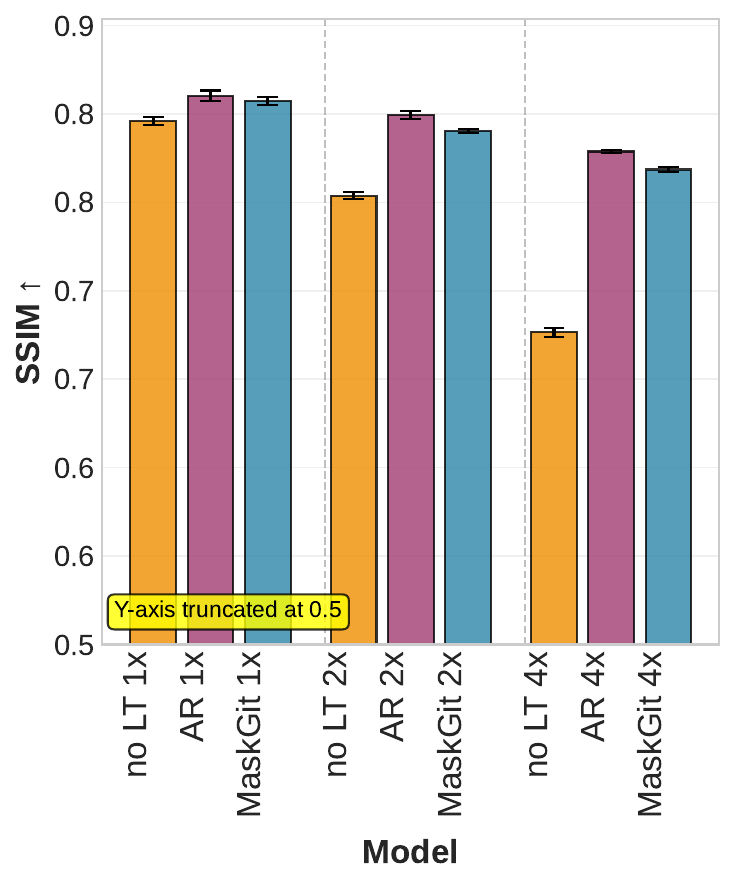}
    \caption{SSIM (seen speakers)}
    \label{fig:1c}
  \end{subfigure}
  \hfill
  \begin{subfigure}{0.24\textwidth}
    \includegraphics[width=\linewidth]{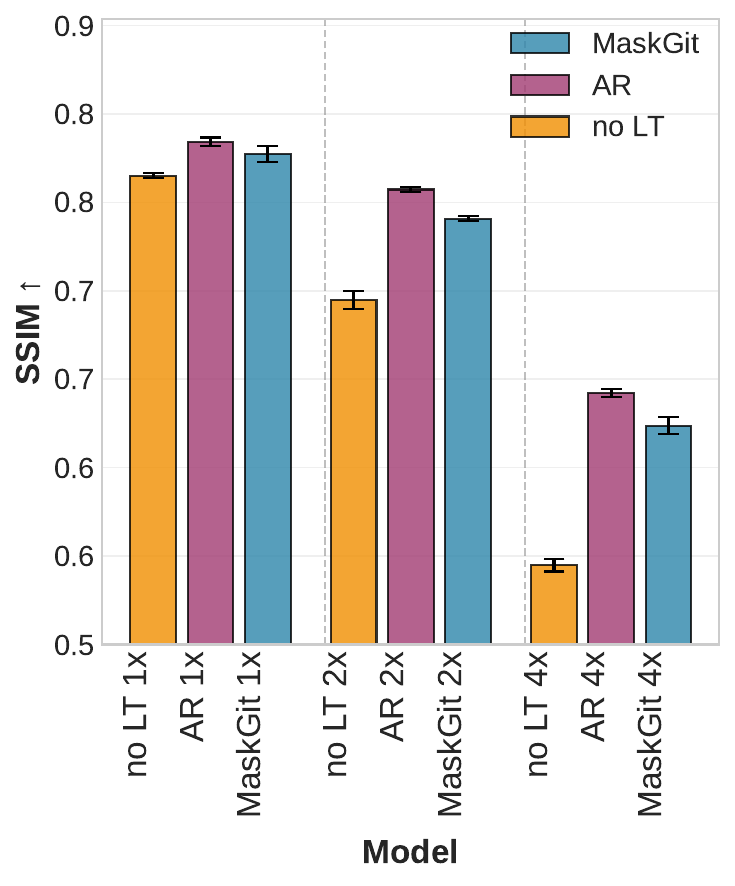}
    \caption{SSIM (unseen speakers)}
    \label{fig:1d}
  \end{subfigure}
~ \\[1.5em]
\noindent 
    \subcaptionbox{Fréchet Distance (unseen speakers)\label{fig:1e}}{%
        \includegraphics[width=0.24\textwidth]{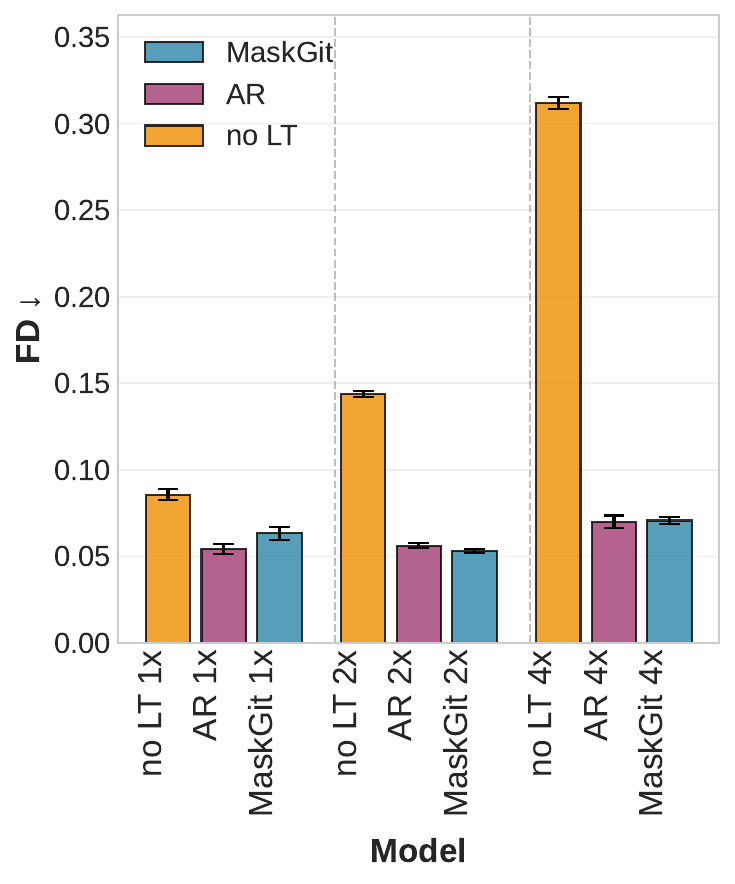}%
    }\hfill
    \subcaptionbox{Inference Speed\label{fig:1f}}{%
        \includegraphics[width=0.24\textwidth]{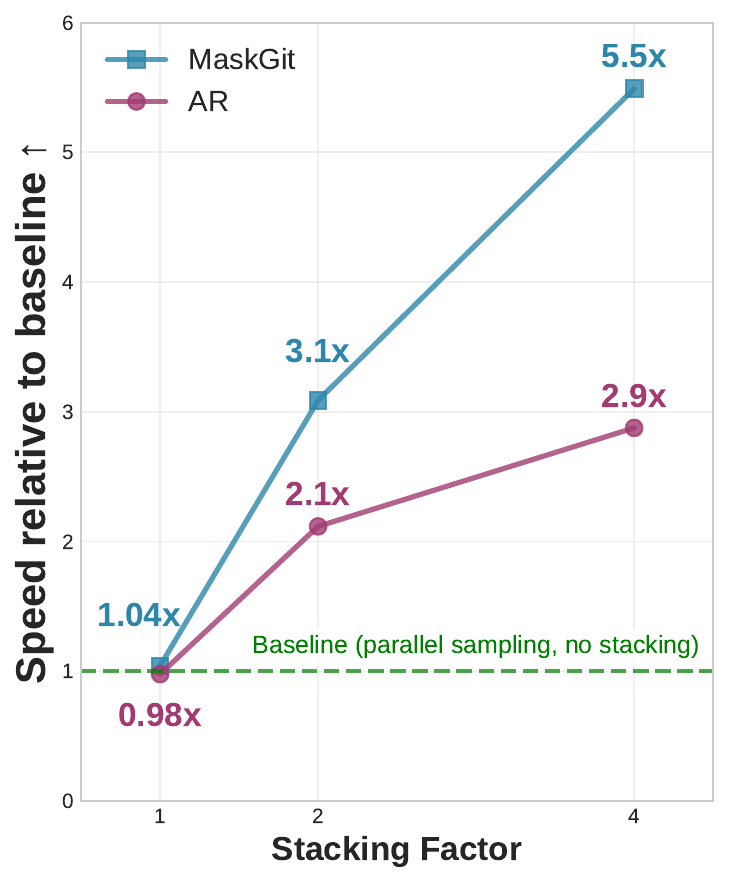}%
    }\hfill
\hfill
    \subcaptionbox{Metrics on LibriTTS. We report 95\% CIs. Bold indicates the best mean score; no bolding is applied if CIs overlap. \label{fig:eval_results}}{%
    \begin{minipage}{0.48\textwidth}
    \resizebox{\linewidth}{!}{%
    \begin{tabular}{>{\centering\arraybackslash}p{1.2cm}>{\centering\arraybackslash}p{0.9cm} c c c c c}

    \toprule
    Eval Set & Stack Factor & LT Type & WER(\%)$\downarrow$ & SSIM$\uparrow$ & FD$\downarrow$ & UTMOSv2 $\uparrow$ \\

    \midrule
\multirow{9}{1.2cm}{\centering\arraybackslash\makecell{Seen\\Speakers}} & \multirow{3}{*}{1} & none &1.1 $\pm$ 0.2&0.796 $\pm$ 0.002&0.089 $\pm$ 0.003&3.54 $\pm$ 0.06  \\
      & & MaskGIT   &1.4 $\pm$ 0.2&0.807 $\pm$ 0.002&0.050 $\pm$ 0.002&3.67 $\pm$ 0.06  \\
      & & AR        &1.2 $\pm$ 0.3&0.810 $\pm$ 0.003&0.049 $\pm$ 0.002&3.66 $\pm$ 0.05  \\
    \cmidrule(lr){2-7}
      & \multirow{3}{*}{2} & none   &1.1 $\pm$ 0.2&0.754 $\pm$ 0.002&0.161 $\pm$ 0.003&3.47 $\pm$ 0.06  \\
      & & MaskGIT   &1.1 $\pm$ 0.3&0.790 $\pm$ 0.001&0.055 $\pm$ 0.001&3.63 $\pm$ 0.05  \\
      &  & AR       &1.1 $\pm$ 0.4&\textbf{0.799} $\pm$ 0.002&0.057 $\pm$ 0.002&3.70 $\pm$ 0.05  \\
     \cmidrule(lr){2-7}
      & \multirow{3}{*}{4} & none  &1.4 $\pm$ 0.2&0.676 $\pm$ 0.003&0.281 $\pm$ 0.004&3.27 $\pm$ 0.06  \\
      & & MaskGIT   &1.1 $\pm$ 0.2&0.769 $\pm$ 0.002&0.061 $\pm$ 0.002&3.45 $\pm$ 0.06  \\
      & & AR        &1.2 $\pm$ 0.1&\textbf{0.779} $\pm$ 0.001&0.060 $\pm$ 0.003&\textbf{3.68} $\pm$ 0.05  \\
      \midrule
      \multirow{9}{1.2cm}{\centering\arraybackslash\makecell{Unseen\\Speakers}} & \multirow{3}{*}{1} & none  &1.2 $\pm$ 0.1&0.765 $\pm$ 0.001&0.086 $\pm$ 0.003&3.57 $\pm$ 0.05  \\
      & & MaskGIT   &1.5 $\pm$ 0.4&0.777 $\pm$ 0.005&0.063 $\pm$ 0.004&3.68 $\pm$ 0.05  \\
      & & AR        &1.3 $\pm$ 0.3&0.784 $\pm$ 0.002&\textbf{0.054} $\pm$ 0.003&3.66 $\pm$ 0.05  \\
      \cmidrule(lr){2-7}
      & \multirow{3}{*}{2} & none  &1.2 $\pm$ 0.1&0.695 $\pm$ 0.005&0.144 $\pm$ 0.002&3.46 $\pm$ 0.06  \\
      & & MaskGIT  &1.3 $\pm$ 0.3&0.741 $\pm$ 0.002&\textbf{0.053} $\pm$ 0.001&3.63 $\pm$ 0.05  \\
      & & AR       &1.0 $\pm$ 0.1&\textbf{0.757} $\pm$ 0.002&0.056 $\pm$ 0.002&3.70 $\pm$ 0.05  \\
      \cmidrule(lr){2-7}
      & \multirow{3}{*}{4} & none &1.5 $\pm$ 0.5&0.545 $\pm$ 0.004&0.312 $\pm$ 0.004&3.22 $\pm$ 0.06  \\
      & & MaskGIT   &1.1 $\pm$ 0.1&0.624 $\pm$ 0.005&0.071 $\pm$ 0.002&3.41 $\pm$ 0.06  \\
      & & AR        &1.1 $\pm$ 0.3&\textbf{0.642} $\pm$ 0.002&0.070 $\pm$ 0.004&\textbf{3.71} $\pm$ 0.05  \\
    \bottomrule
  \end{tabular}
}
\end{minipage}
}
\caption{\small{Evaluation Results on LibriTTS. 1x, 2x, 4x in the model names refer to the number of frames stacked. (a) LT models achieve same or better UTMOSv2 scores than baseline, except the 4x-stacked MaskGit LT. (b) WERs are similar for all models, with CIs overlapping. (c)(d) SSIMs: at 1x stacking, LT models have an advantage; at 2x, LT models are similar to baseline; at 4x, LT models are still usable for seen speakers, but with reduced robustness to unseen speakers. (e) FDs: LT models consistently exhibit a strong advantage over parallel sampling. (f) Significant inference speedup from frame stacking. (g) Metrics table.}}
\vspace{-5mm}
\end{figure*}

\subsection{Training and Sampling Setup}
We base our experiments on the TTS model described in Koel-TTS \cite{hussain2025koel}. To keep the number of transformer layers fixed between the baseline and LT models, we add 4 transformer layers to the baseline's decoder such that it has 16 layers while the LT-based models have 12 in main decoder and 4 in the LT, for the same total of 16 layers. All transformer decoder layers (primary and LT) have dimension 768 and 12 self-attention heads. We experiment with stacking factors of 1 (no stacking), 2 and 4. We train all models on the same 18k hours of data as in the Koel-TTS paper for 220k steps with the AdamW~\cite{loshchilov2018decoupled}. For sampling, we use the same CFG, top-k, and temperature setup as \cite{hussain2025koel}. For models with a MaskGIT-based LT we use 3 sampling steps and choose which tokens to unmask using \textit{purity sampling}~\cite{tang2022improved,lee2023text}.
\vspace{-2ex} 
\subsection{Evaluation Criteria}
We evaluate the model on the same seen and unseen-speaker subsets of LibriTTS \cite{libritts} as \cite{hussain2025koel}, containing 180 utterances each. To assess text adherence, we report the Word Error Rate (\textbf{WER}) computed with the Parakeet-TDT-1.1b ASR model \cite{rekesh2023fast}. For \emph{Speaker Similarity} (\textbf{SSIM}) , we extract speaker embeddings using TitaNet-Large \cite{koluguri2022titanet} and compute the cosine similarity between the generated and reference (context) audio. Following \cite{pascual2024masked} and others, we compute a Fréchet Distance (\textbf{FD}) in the codec's 32-dimensional embedding space. We use model-generated frames as the generated distribution and the ground truth LibriTTS codec frames of the same utterances as the real distribution. This metric measures how closely the distribution of generated and real codec frames match, capturing both fidelity and diversity. If a sampling method yields implausible codebook token combinations, this metric should detect it. We report speech quality using \textbf{UTMOSv2} \cite{baba2024utmosv2}, a state-of-the-art neural estimator of naturalness mean opinion scores (MOS). Inference speed (i.e., throughput) is reported relative to the baseline's speed.\footnote{To ensure our quality comparisons were fair, the baseline model was designed with 16 decoder layers to match the total layer count (12+4) of the LT models. While this establishes an equitable footing for quality metrics, it slightly inflates speedup figures. A baseline with only 12 layers would be roughly 14\% faster, which would proportionally lower the speedup figures we report. Nevertheless, the fundamental conclusion that frame-stacked models offer significant throughput gains remains valid.
}

\subsection{Results}
Table \ref{fig:eval_results} provides a full set of evaluation metrics on LibriTTS.
In all reported results, \emph{baseline} denotes the non-stacked model (1x-stacked) with no LT, parallel-sampled across all codebooks of a single frame. Speeds are reported in Figure \ref{fig:1f}. We next review these results. 
\vspace{-1.5ex} 
\subsubsection{Iterative Sampling and FD}
Figure \ref{fig:1e} shows the FDs for unseen speakers (trends for seen speakers are similar). At each stacking factor, the LT-based models have a lower (better) FD than the no-LT (parallel-sampled) model. In fact, remarkably, all LT-based models, at all stacking factors, have a lower FD than all parallel models, even the unstacked one. This result supports the hypothesis that iterative sampling generates a distribution that is closer to the ground truth than parallel sampling.
\vspace{-1.5ex} 
\subsubsection{Local Transformer and Frame Stacking}
\textbf{At a frame stacking factor of 1}, SSIMs and MOS estimates for both LT-based models outperform baseline for both seen and unseen speakers. WERs are slightly higher for LT models compared to baseline, but the difference is not statistically significant. Speeds are similar across the three models. Overall, the LT-based models are a better choice at this stacking factor, producing better SSIM, MOS and FD at similar speed.
\textbf{At a frame stacking factor of 2}, we start observing major speed benefits from frame stacking. The AR LT is \textbf{2.1x} faster and the MaskGIT LT \textbf{3.1x} faster than the unstacked parallel baseline. SSIMs are similar to baseline for seen speakers and nearly as good for unseen speakers. MOS scores are better or within CI of baseline. If we try to \emph{parallel}-sample from a 2x-stacked model, the sampling starts to degrade substantially, with FDs in particular worsening by 67\% for unseen speakers and MOS decreasing. This is not surprising as parallel sampling two frames, rather than one, can only exacerbate the issues with parallel sampling discussed earlier. Overall, even at a stacking factor of 2, the LT-based models are a better choice than the baseline. \textbf{At a frame stacking factor of 4,} there are large speedups of 2.9x (LT) and 5.5x (MaskGIT) vs baseline but with some cost in quality and robustness: SSIMs drop a little for seen speakers but substantially for unseen speakers. WER confidence intervals remain overlapping with baseline. FDs remain better than baseline. MOS for the AR LT remains similar to baseline but for MaskGit there is a significant drop. We attribute this drop to our use of only 3 sampling steps—a constraint likely too severe for sampling $8\times 4=32 $ tokens—and expect that relaxing it could mitigate the degradation. Parallel sampling breaks at this stacking factor, with large large degradations in both FD and SSIM. Based on these results, \textbf{we propose the following practical guidelines:}
\begin{itemize}[topsep=0pt, partopsep=0pt, itemsep=2pt, parsep=0pt]

    \item When quality is the primary consideration, the non-frame-stacked configuration with an autoregressive local transformer is recommended.
    \item A good balance between quality and complexity can be achieved by using frame stacking factor of 2 with an autoregressive LT. This works as well as baseline but is 2.1x faster. MaskGIT also achieves good performance at a 3.1x speedup.
     \item When aggressively seeking speedup, and not needing zero-shot functionality, use a high stacking factor (e.g. 4) with a LT (either AR or MaskGIT).
    
\end{itemize}
\vspace{-2ex} 
\section{Conclusion}
\label{sec:conclusion}
We study iterative multi-codebook prediction methods using Local Transformers for TTS models and provide a comparative analysis against parallel prediction. We demonstrate that iterative prediction is important to capture intra-codebook dependencies, improving audio fidelity. LTs also enable frame-stacking in the primary decoder, which substantially improves the throughput without having to retrain a lower frame rate codec model.


\bibliographystyle{IEEEbib}
\bibliography{refs}
\end{document}